\documentclass[amsmath,amssymb,aps,pra,twocolumn]{revtex4-2}
\usepackage{amsmath,amsthm,amssymb,mathtools,braket,hyperref}
\usepackage[T1]{fontenc}
\usepackage{xcolor} 

\newtheorem{theorem}{Theorem}
\newtheorem{proposition}{Proposition}
\newtheorem{lemma}{Lemma}
\newtheorem{corollary}{Corollary}
\newtheorem{definition}{Definition}

\newcommand{\Hilb}{\mathcal H}
\newcommand{\Herm}{\operatorname{Herm}}
\newcommand{\Eff}{\mathcal E}
\newcommand{\Acts}{\mathcal F}
\newcommand{\Dens}{\mathcal D}
\newcommand{\Tr}{\operatorname{Tr}}
\newcommand{\aff}{\operatorname{aff}}

\newcommand{\Rmax}{\mathcal R_{\max}}
\newcommand{\Pcal}{\mathcal P}

\newcommand{\Ccal}{\mathcal C}

\begin{document}

\title{An Incomplete-Preference Quantum Anscombe--Aumann Representation}

\author{Cristhiano Duarte}
\email{cristhianoduarte@gmail.com}
\affiliation{Universidade Federal de Juiz de Fora, Departamento de Física, Juiz de Fora, MG, Brazil}
\affiliation{Institute for Quantum Studies, Chapman University, One University Drive, Orange, CA, 92866, USA}
\affiliation{Fundação Maurício Grabois, R. Rego Freitas, 192 - República, São Paulo - SP, 01220-010, Brazil}
\date{\today}

\begin{abstract}
By dropping the idea that completeness is a mandatory prerequisite for agents' rationality, this work shows how classical and quantum subjective probability attributions represent suitable preference orderings. To do so, we develop an extension of the Anscombe--Aumann framework in which acts map a finite set of states of the world into the effect interval of a finite-dimensional Hilbert space, while preferences need not be complete. The construction keeps three logically distinct questions separate. First, an incomplete-preference multi-utility theorem supplies affine utility functionals. Second, finite-dimensional operator geometry determines the form of each affine functional. Third, admissibility identifies which normalised probability--state pairs are compatible with the full preference. We introduce behavioural state-independence as the natural Anscombe--Aumann interpretation of the idea that the identity of the states of the world should not affect how quantum consequences are compared. Because an incomplete preference makes this aggregate behavioural condition weaker than a utility-by-utility condition, the factorisation theorem requires a stronger, explicitly operational representation-level condition. Under that condition and Loewner monotonicity, every affine utility functional is positively affinely equivalent to a normalised utility of the form $U_{p,\rho}(f)=\sum_s p(s)\Tr(\rho f(s))$. Consequently, a suitable incomplete preference relation is represented by a family of admissible pairs in $\Delta(S)\times\mathcal D(\Hilb)$. We then identify a canonical maximal compatible family and show that the scalar-effect subspace carries an ordinary classical multi-utility representation, with scalar events appearing as the vertices of the scalar-act cube.
\end{abstract}
\maketitle

\section{Introduction}\label{Sec.Introduction}

\emph{Completeness}, or the idea that rational agents know exactly how to rank each and every action they can make, in every possible state of the world, is highly artificial and has been seriously criticised. In the words of Robert Aumann himself: \emph{``of all the axioms of utility theory, the completeness axiom is perhaps the most questionable. Like other axioms, it is inaccurate as a description of real life; but unlike them, we find it hard to accept even from the normative viewpoint''} \cite{Aumann62}. So, perhaps, what we should do is to abandon it altogether. However, this alternative is not that simple.

Trying to remove completeness from the picture opens up a roadmap of possibilities \cite{Aumann62, AA63, Neurath1983, Benanav20, Fritz17,GK12}. Not least, we lose the inherent ability to compare actions with a single (affine-)utility function \cite{GK12,McCMT21,Fritz17}---and for operational theories with a physical significance, losing a single entropy-like function may not be what we set out to obtain in the first place \cite{LY99}. Therefore, despite its artificiality, one should be cautious, well-justified and attentive to the effects of assuming that agents have only an incomplete preference relation over their actions.  Abandoning completeness, with an operational justification of why we are doing so, and carefully analysing its consequences is exactly what we propose to do in this work.

Inspired by Anscombe's and Aumann's argument for the existence of a classical, subjective probability associated with naturally uncertain events \cite{AA63}, we investigate what potentially incomplete preference relations over the collection of all acts can generate. We argue that, under suitable Anscombe--Aumann-style behavioural hypotheses, it is possible to recover a representation theorem for an incomplete preference relation via an, in general, uncountable family of affine utility functions---see refs. \cite{Turgut07,klimesh07,Datta24, AI08} for a similar situation involving entanglement transformation and its description via infinitely many Rényi entropies. Moreover, in line with the original framework, our setup provides not only the assignment of a classical subjective probability, but also a subjective density operator and a factorisation between those objects in a well-known way. Within this factorisation, one can readily identify an expression resembling the celebrated Born rule \cite{Busch03,BFZ17}, potentially granting it a different, entropy-like, interpretation. 

This contribution should be seen against the background of a larger program that views quantum theory as nothing more than a particular theory of probabilistic assignments---see refs. \cite{Duarte20,BRD25,RodriguesEtAl26,FRS26,LD26, BFZ16, BFZ17}. However, there is a caveat. In parallel with Anscombe and Aumann's original formulation, the framework we explore here also derives subjective probabilistic assignments from physical probabilities. While in Anscombe and Aumann's argument, subjective assignments are constructed on top of classical lottery games \cite{AA63}, here we build our framework assuming that effects are responsible for giving rise to physically motivated probabilities---objects that could be used to empirically determine the frequency of occurrence of a physical event in an experiment. In other words, we will somewhat keep Anscombe and Aumann's neat distinction between chances (lottery games or physical probabilities) and degrees of belief (horse races or probability proper). Our main result is that by granting rational agents a suitable preference relation, we can $(i)$ work out how this preference relation can be faithfully expressed by a family of (affine-)utility functionals, and $(ii)$ explore under which circumstances this family splits into an expression resembling the well-known Born rule. 

To do so, this work is organised as follows. In section \ref{Sec.AAModel} we briefly review the original Anscombe and Aumann's model and discuss how our own framework will differ from it. We go into more detail about the (suitable) preference relation we will use throughout the paper in section \ref{SubSec.FrameworkBehaviouralInput}. We take a small mathematical detour in section \ref{Sec.AffineGeometryOperatorRepresentation}, and get back on track in secs. \ref{Sec.ProportionalityFactorisation} and \ref{Sec.PositivityAndDensityNormalisation} to show when one can rewrite the representing affine functions in a physically meaningful way.  Subsections \ref{SubSec.ClassicalSubspace} and \ref{SubSec.StateContigentQuantumSlices} focus on the classical and quantum slices of our representing family. We analyse our work and contrast it with others in the field in the last section \ref{Sec.DiscussionConclusion}.

\section{Anscombe-Aumann Model}\label{Sec.AAModel}

\subsection{Classical Model - Subjective Utility and Probability Assignment}\label{SubSec.ClassicalAAModel}

The classical Anscombe--Aumann framework \cite{AA63}  has a simple conceptual architecture: a finite set of \emph{states of the world} $S$; a usually finite \emph{prize set} $Z$; a \textit{consequence set}
\begin{equation}
X=\left\lbrace x:Z \rightarrow [0,1]; \sum_{z \in Z}x(z) =1 \right\rbrace,
\label{Eq.DefClassicalConsequences}
\end{equation}
where each element is a random variable with an exogenously determined probability distribution; and a \textit{set of actions} 
\begin{equation}
\mathcal{A}=X^{S}:=\{f: S \rightarrow X\},
\label{Eq.DefClassicalActs}
\end{equation}
connecting each state of the world to a random variable in $X$. Over $\mathcal{A}$ it is imposed a \textit{suitable preference relation}
\begin{align}
    \succeq \,\, \subseteq \mathcal{A} \times \mathcal{A}
\end{align}
capturing each agent's preference relation and satisfying (for every $f$, $g$, and $h$ in $\mathcal{A}$):
\begin{itemize}
    \item[(i)] $f \succeq g$ or $g \succeq f$ \emph{(completeness)};
    \item[(ii)] if $f \succeq g$ and $g \succeq h$, then $f \succeq h$ \emph{(transitivity)};
    \item[(iii)] there exists at least a pair $(h',h'') \in \mathcal{A} \times \mathcal{A}$ with $h' \neq h''$ such that $h' \succeq h''$ \emph{(non-triviality)};
    \item[(iv)] that the sets $\{\alpha \in [0,1]; \alpha f + (1-\alpha) g \succeq h \}$ and $\{\alpha \in [0,1]; h \succeq \alpha f + (1-\alpha) g \}$ are closed \emph{(continuity)};
    \item[(v)] for every $\alpha \in (0,1)$, $f \succeq g \Leftrightarrow \alpha f + (1-\alpha) h \succeq \alpha g + (1-\alpha) h$  \emph{(independence)};
    \item[(vi)] if $f(s) \succeq_X g(s)$ for every $s$ in $S$, then $f \succeq g$ \emph{(monotonicity)}.
\end{itemize}
\textbf{Remark:} In item $(vi)$ above, $\succeq_X \ \subseteq X \times X$ denotes the consequence ordering induced by $\succeq$: for $x,y\in X$, write $x\succeq_X y$ whenever the constant acts $C_{x}, C_{y} \in \mathcal{A}$, defined by $C_{x}(s)=x$ and $C_{y}(s)=y$ for every $s\in S$, satisfy $ C_{x} \succeq C_{y}$.

The heart of the Anscombe-Aumann setup is that this suitable preference relation can be conveniently expressed as a comparison determined by an affine functional---see Thm.\ref {Thm.ClassicalAA} below. We will get there in a minute, but before that, a few preliminary observations are in place. 

First, following von Neumann--Morgenstern's decision-theoretic framework \cite{vNM44}, Anscombe and Aumann call elements $x$ in $X$ of (roulette) lotteries and the acts $f$ in $\mathcal{A}$ of horse races. This distinction echoes a strand in the foundations of probability that emphasises the potential distinction between physical probabilities (naturally occurring chances) and logical probabilities or degrees of belief (a person's description of unknown phenomena) \cite{Gillies10}. Second, although Anscombe and Aumann do not explicitly mention that in their work, the idea of defining a subjective probability from an agent's action can be traced back to Ramsey's own original take on the subject, as he emphasised that the kind of measurement of belief with which probability is concerned is a measurement of belief qua basis of action---see \cite{Ramsey1931}. Third, of all properties imposed on the preference relation, completeness and monotonicity are particularly remarkable: \emph{(a)} as we mentioned in the introduction, completeness is rather artificial, and it is by carefully dropping it that we get a richer result; \emph{(b)} monotonicity acts like a bridge and says that a state-by-state preference extends to a global preference, and it is exactly at this point that one may use von Neumann--Morgenstern kind of argument to pass on and derive Anscombe and Aumann's result: a representation theorem for a preference $\succeq \,\, \subseteq \mathcal{A} \times \mathcal{A}$ in terms of a utility function \cite{AA63,CF05,Kreps1988}.

\begin{theorem}[Classical Anscombe-Aumann]
    Let $\succeq$ be a preference relation over $\mathcal{A}$ satisfying \emph{(i)}-\emph{(vi)} above. Then, there exists a unique probability $p:2^{S} \rightarrow [0,1]$ and a utility function $u:X \rightarrow \mathbb{R}$ such that for every $f$ and $g$ in $\mathcal{A}:$
    \begin{align}
        f \succeq g \Leftrightarrow \sum_{s \in S} u(f(s))p(s) \geq \sum_{s \in S} u(g(s))p(s)
        \label{Eq.ThmClassicalAA}
    \end{align}
    Moreover, $u$ is unique up to an affine transformation.
    \label{Thm.ClassicalAA}
\end{theorem}

There is an implicit consequence in the Anscombe--Aumann representation result. Beyond the existence of a subjective probability, it says that suitable preference relations over a set of actions can be equivalently determined by a numerical evaluation. This is far from trivial. It is as if all the richness of a rather abstract ordering relation could be compressed and expressed by the single comparison of two real numbers. In physics, this reading of Anscombe's and Aumann's result is not entirely a novelty. Textbook thermodynamics can be viewed as a resource theory adapted to investigate which configurations are accessible from the others (a suitable preference relation) and, not so surprisingly, that accessibility relation is neatly expressed through a single entropy (a single utility function)---see refs. \cite{LY99,Fritz17,Callen85}. Obviously, when completeness is dropped out, the situation is far more complex \cite{GK12}, but the entropy-like reading remains. For instance, in quantum theory, under the local operational and classical communication paradigm, the possibility of accessing an entangled state from another can be promptly read off from the comparison of an infinitely large set of (Rényi) entropies \cite{klimesh07,BFZ16,Nielsen99}.   

\subsection{Classical AA Model - A Historical Note}\label{SubSec.ClassicalAAModelHistoricalNote}

We conclude this brief overview with a historical reconstruction. Presumably, one of the most well-known justifications for a subjective expected utility theory was provided by Savage \cite{Savage54}, following a particular scholarship in decision theory and foundations of probability previously initiated by Ramsey \cite{Ramsey1931}, von Neumann \cite{vNM44} and de Finetti \cite{deFinetti37}---and later also advocated by Jeffrey \cite{Jeffrey1965}. Savage's hallmark contribution was to derive both subjective utilities and probabilities from agents' preference relations. His axioms are well-motivated, but the derivation itself is rather convoluted and does not avoid criticism, as it requires a rich set of available acts and states of the world. Soon after the publication of Savage's seminal work, several authors attempted to simplify his derivation: Rubin \cite{Rubin87}, Blackwell and Girshick \cite{Blackwell54}, Chernoff \cite{Chernoff54}, Pratt, Raiffa and Schlaifer \cite{PRS64} and, in particular,  Anscombe and Aumann \cite{AA63}. Broadly speaking, by maintaining a distinction between chances and degrees of belief, those simplifications were obtained by introducing an objective and external randomising device, such as a roulette wheel, on top of which the subjective utility and probability functions would be obtained. These alternative derivations, therefore, used a two-stage setup in which an act yields an objective probability distribution (a roulette wheel) for each state of the world, and subjective probabilistic assignments follow from it. Although slightly artificial, the two-stage setup pays dividends by simplifying the set of states of the world, the hypothesis on the preference order over the acts and, consequently, the proofs.

Perhaps because it works for finite state spaces and because the demonstrations are far less complicated than those of its rivals \cite{Kreps1988}, Anscombe and Aumann's approach has been considered the most popular alternative to Savage's subjective expected utility derivation. As we mentioned before, they borrow the monotonicity axiom of von Neumann and Morgenstern \cite{vNM44} for the two-stage setup by means of a reversal of order in the compound lotteries assumption \cite{AA63}. Their original paper may not be reader-friendly, but their derivation is certainly much simpler than Savage's \textit{tour de force}.   

Finally, as we set out to do in this contribution, several authors have fruitfully employed the Anscombe-Aumann setup to refine and modify the subjective expected utility model---see refs. \cite{Fishburn82, Hazen87, Schmeidler1989, Karni93,Nau06,EK96,machina1995bayes}. Although we are also adding to this already prolific discussion, the main difference is that by dropping completeness and generalising the idea of physical probabilities (chances) to quantum effects, we steer the discussion towards the foundations of quantum theory and obtain a subjective assignment of a density operator, together with a Born-rule-like utility function. 

We should emphasise, though, that we are not the first to chart these waters. There have also been quantum-like alternatives to decision-theoretical frameworks in the recent past---see mainly the refs.\cite{BFZ17, PieraEtAl25} and references therein. Distancing from the bet-setup of those works and abandoning the principle of sure loss as a rationality criterion \cite{Pettigrew20}---which, in itself, already carries a certain idea of commensurability that we want to purge from defining agents' preference relations \cite{Neurath1983}---the main difference is that here we are working directly at the agent's preference level and examining which subjective, preference-based elements contribute to the emergence of a family of utility functions. We will return to this discussion later in this work. For now, let us pass on to our own model. 

\section{Quantum Anscombe-Aumann model}\label{Sec.QuantumAAModel}

\subsection{Introduction}\label{SubSec.QuantumAAModelIntro}

To obtain a subjective, quantum-like probabilistic assignment, the approach we develop here combines elements of the standard Anscombe-Aumann framework with those of quantum theory. From now on, then, let $S$ be a finite set representing the possible states of the world, and let $\Hilb$ be a finite-dimensional Hilbert space. Borrowing from Anscombe and Aumann, our consequence and act sets are defined as follows.

\begin{definition}[Consequence Set]
    The consequence set is the effect interval
    \begin{align}
        \Eff=\{E\in\Herm(\Hilb):0\preceq E\preceq I\},
        \label{Eq.DefQuantumConsequenceSpace}
    \end{align}
    where $I$ is the identity over $\Hilb$, and $E \preceq I$ means that $A:=I-E$ is a positive, semidefinite operator in $\Herm(\Hilb)$. 
    \label{Def.QuantumConsequenceSpace}
\end{definition}

\begin{definition}[Act Set]
    An act is simply a function from the set of states of the world to the effect interval, more explicitly:
    \begin{align}
        f:S &\to \Eff \nonumber \\
          s &\mapsto 0 \preceq f(s) \preceq I.
    \end{align}
    The set of all quantum-like acts is denoted by $\mathcal{F} = \Eff^{S}$.
    \label{Def.QuantumActSet}
\end{definition}

In line with Anscombe and Aumann \cite{AA63}, an act may be interpreted as a classical contingent plan whose consequences trigger a quantum effect. In other words, to each state of the world that plan assigns one and only one effect, as if it showed what course of action an experimentalist would take in their laboratory should a particular state of the world transpire. To create an even closer parallel with Eq. \ref{Eq.DefClassicalConsequences} we might have defined each act as a POVM by requiring that $\sum_{s \in S} f(s) = I$; nonetheless, we have refrained from doing so for several reasons. Not least because our framework is more general, but also because it would beg the question and require another layer to justify why additivity holds. 

Again, in this work, we address two main questions. First, we investigate under which conditions there exists a set of utility functions representing a suitable preference order over $\mathcal{F}$. Secondly, when such a utility exists, we study whether affine utility evaluations of acts in $\mathcal{F}$ separate into a probability over classical states and a normalised quantum functional on effects. Combining both objectives, we want to know whether one can represent a suitable preference relation by a family of utility functions as follows: 
\begin{align}
    U_{p,\rho}(f)=\sum_{s\in S}p(s)\Tr(\rho f(s)),
    \label{Eq.DesiredFormForQuantumUtility}
\end{align}
where $p$ is a probability distribution in the simplex $\Delta(S)=\{p:S \to [0,1]; \sum_{s \in S}p(s)=1\}$ and $\rho$ is a density operator in $\mathcal{D}(\Hilb)=\{\rho \in \Herm(\Hilb); \rho \succeq 0 \mbox{ and } Tr(\rho)=1 \}$. As we will see in a minute, the answer is affirmative under a set of structural restrictions---at the level of the preference $\succeq$ over $\Acts$ as well as at the operational level of the utilities themselves. When this is the case, the two objects in Eq.\eqref{Eq.DesiredFormForQuantumUtility} play different roles. While $p \in \Delta(S)$ describes how the classical states are weighted, $\rho \in \mathcal{D}(\Hilb)$ describes how quantum consequences ought to be evaluated. The representation, therefore, preserves the central Anscombe--Aumann separation even though the consequence space is no longer an interval of real numbers.

The incomplete-preference setting is important. We do not assume that every pair of acts is comparable. Consequently, there need not be a single pair $(p,\rho)$ representing the preference. Instead, as we will see, the preference is fully captured by an indexed family $\{ U_{p,\rho}\}_{(p,\rho) \in \mathcal{R}}$, where:
\begin{align}
\mathcal{R}\subseteq\Delta(S)\times\mathcal D(\Hilb).
\end{align}
is the admissible set---see definition \ref{Def.AdmissibleFamily}.

\subsection{Informal Statement, Sketch of the Proof and Comments on the Result}\label{SubSec.InformalQAAStatement}

We argue that an act $f \in \mathcal{F}$ is (suitably) preferred to another $g \in \mathcal{F}$ precisely when every utility function $U_{p,\rho}$ whose indices belong to the admissible set $\mathcal{R}$ agrees with that comparison.

\begin{theorem}[Informal Version]
    If $\succeq$ is a suitable preference relation over $\mathcal{F}$, and a chosen affine multi-representing family satisfies representation-level state-independence and Loewner monotonicity, then for every pair of acts $f,g \in \mathcal{F}$:
\begin{align}
    f \succeq g &\iff U_{p,\rho}(f) \geq U_{p,\rho}(g), \forall \,\, (p,\rho) \in \mathcal{R}.
    \label{Eq.InformalExpressionOfMainThm}
\end{align}
\end{theorem}
The fact that we have to deal with an indexed family is not an incidental technical device: it is the natural representation of each agent's incomplete preference relation expressed as an unresolved uncertainty about the relevant probability-quantum-state pair in the admissible set $\mathcal{R}$. As will become clearer later, we had to add representation-level state-independence and Loewner-monotonicity to the mix, so we can split the utility function into classical and quantum contributions. We are not claiming that these two are the minimal assumptions for that derivation, only that we had to impose extra conditions that have no parallel in the original Anscombe--Aumann framework. 

The proof is organised in three layers. First, an external finite-dimensional multi-utility theorem supplies affine supporting functionals without imposing completeness \cite{McCMT21}. Second, affine geometry and the Hilbert--Schmidt pairing turn those functionals into statewise Hermitian coefficient operators \cite{HK71}. Third, an explicit state-independence restriction forces those operators to factor into a classical weight and a common positive operator, which can then be normalised to a density operator. This onion structure is rather illuminating, as it allows us to keep the operator conclusions at bay and away from the behavioural representation theorem (see Thm \ref{Thm.MultiValuedMixtureRep})---recall that we are addressing two main questions, the first about the possibility of a faithful representation and a second about its form.  

There is one subtlety specific to incompleteness. Behaviourally, at the level of the preference order $\mathcal{\succeq}$ over acts in $\mathcal{F}$, it is natural to require that the comparison of two quantum consequences not depend on which classical state carries them. We call this sort of non-contextual behavioural choice \emph{state-independence}. For complete preferences, it is easy to overlook the distinction between this statement and the corresponding condition on a representing utility. With incomplete preferences, however, the aggregate preference (the r.h.s in Eq. \eqref{Eq.InformalExpressionOfMainThm}) is an intersection of the orderings represented by its associated utilities. The behavioural condition, therefore, does not automatically imply that every individual utility has the same statewise ordering as it should. We make this distinction explicit and use the stronger representation-level condition only where the factorisation proof requires it.

Finally, we circle back to the classical-like part of the model. The scalar-effect acts, $f_x(s)=x(s)I$, form a genuine copy of the classical mixture space $[0,1]^S$. On this subspace, as one may anticipate, the density operator drops out of the evaluation, leaving only $p$. Deterministic events are the vertices of this subspace, and the induced event comparisons therefore recover upper and lower probability information from the projected admissible family. All things considered, our result is less a replacement for classical Anscombe--Aumann theory than a controlled extension of its architecture: classical uncertainty remains classical, quantum-like consequences are evaluated using density operators, and incompleteness is represented by a family rather than a single pair.

\subsection{Framework and behavioural input}\label{SubSec.FrameworkBehaviouralInput}

The centre of this work is a definition of a suitable preference relation over acts that allows for a representation in terms of affine utility functions as in Eq.\eqref{Eq.InformalExpressionOfMainThm}. Because they are meant to express agents' subjective reasoning, we refer to conditions imposed over $\succeq \ \subseteq \mathcal{F} \times \mathcal{F}$ as behavioural inputs. We start from the preference relation itself.

\begin{definition}[Preference Relation]
A preference over $\mathcal{F}$ is a binary relation 
\begin{align}
\succeq \,\, \subseteq \mathcal{F} \times \mathcal{F}
\end{align}
such that for all $f,g,h \in \mathcal{F}$:
    \begin{enumerate}
        \item [(i)] $f \succeq f$ (reflexivity).
        \item [(ii)] If $f \succeq g$ and $g \succeq h$, then $f \succeq h$ (transitiveness).
        \item [(iii)] there exists at least one pair $(f,g)$ of acts in $\mathcal{F}$ such that $f \succeq g$ and $g \nsucceq f$ (non-triviality).
    \end{enumerate}
    \label{Def.IncompletePreference}
\label{Def.PreferenceRelationOverF}
\end{definition}

A central aspect of the present work is that completeness is not assumed. We do not impose that for every pair of acts $f,g \in \mathcal{F}$ either 
$f \succeq g \mbox{ or } g \succeq f$. Nonetheless, borrowing from Anscombe-Aumann's original argument, we do impose a form of strong independence: 

\begin{definition}[Strong Independence]
For all $f,g,h \in \Acts$ and $\lambda\in(0,1)$,
\begin{align}
f\succeq g\iff \lambda f+(1-\lambda)h\succeq\lambda g+(1-\lambda)h.
\end{align}
    \label{Def.StrongIndependence}
\end{definition}

We also demand the mixture-continuity condition, as required by the incomplete-preference multi-representation theorem (Thm.\ref{Thm.MultiValuedMixtureRep}) used below.

\begin{definition}[Mixture-Continuity]
    A preference relation $\succeq$ on $\mathcal F$ satisfies \emph{mixture-continuity} if, for every $f,g,h\in\mathcal F$, the sets
    \begin{align}
    \overline{K}=\{\lambda\in[0,1]; \ & \lambda f+(1-\lambda)g\succeq h\} \nonumber \\
    &\mbox{ and } \\
    \underline{K}=\{\lambda\in[0,1]; \ & h\succeq\lambda f+(1-\lambda)g\} \nonumber 
    \end{align}
    are closed in $[0,1]$.
    \label{Def.MixtureContinuity}
\end{definition}

For mixture preorders, this is equivalent to the mixture-continuity axiom used in \cite{McCMT21}. Mixture-continuity then says that $\succeq$ is well-behaved when taking appropriate limits. Together, strong independence (Def.\ref{Def.StrongIndependence}) and mixture-continuity (Def.\ref{Def.MixtureContinuity}) are fair hypotheses if we recall that our goal is to represent a preference order via affine utility functions—the preference order should reflect somehow the fact that in the end we are comparing acts with continuous affine functionals.  


To reflect the conditions imposed on a preference relation without having to repeat them throughout the paper, we give the following definition.

\begin{definition}[Suitable Preference]
A binary relation $\succeq \ \subseteq \Acts \times \Acts$ is a \emph{suitable preference} whenever it is reflexive, transitive, non-trivial, satisfies strong independence, and satisfies mixture-continuity.
    \label{Def.Suitable}
\end{definition}

To faithfully represent a suitable preference order, we will borrow the finite-dimensional version of the expected multi-utility theorem from ref.\cite{McCMT21}. Our scenario matches their hypothesis perfectly, as our act space $\Acts=\Eff^S$ is a convex subset of the finite-dimensional real vector space $\Herm(\Hilb)^S$, whose dimension is $|S|d^2$ when $d=\dim\Hilb$, and that $\aff(\Acts) = \Herm(\Hilb)^S$ up to an isomorphism---see \cite[Proposition 4.1 and Theorem 2.3]{McCMT21}. So that, in their language, the dimension of $\Acts$ is countable and the following representation theorem holds true: 

\begin{theorem}[Multi-Utility Representation Theorem]
    If $\succeq$ is a suitable preference, then there exists a nonempty indexed family $\{U_{\alpha}\}_{\alpha\in L}$ of affine functions $U_{\alpha}:\mathcal{F}\to\mathbb{R}$ such that
    \begin{align}
    f\succeq g\iff U_{\alpha}(f)\ge U_{\alpha}(g)\quad\text{for every }\alpha\in L.
    \end{align}
    \label{Thm.MultiValuedMixtureRep}
\end{theorem}

We will not re-prove this multi-utility representation theorem here. Although not extremely complicated, its demonstration is rather long and would likely throw the reader off scent because of the several details involved in the proof---the reader may want to check ref.\cite{McCMT21}. What is fundamental is that the representation family is non-empty, and each of its members preserves mixtures, and since the act space is convex, these mixture-preserving functions are precisely affine functions.

That said, recall that this only addresses half of our questions. Because we want to signal out density operators as subjective probabilistic assignments, on par with classical, subjective probabilistic assignments—or with ``horse lotteries'', to stick with the nomenclature used in ref.\cite{AA63}—we will impose yet another restriction on our suitable preference relation. Also inspired by Anscombe-Aumann's framework, we will use the following behavioural principle as the natural interpretation of state-independence. 
\begin{definition}[Behavioural state-independence]
For $s\in S$ and $E\in \Eff$, let $f_{s,E}$ denote the act that yields $E$ at $s$ and $0$ elsewhere. In other words, 
\begin{align}
    f_{s,E}: S &\longrightarrow \Eff \nonumber \\
    t &\mapsto f_{s,E}(t)= \begin{cases}
        E, \mbox{ if } s = t \\
        0, \mbox{ if }  s \neq t
    \end{cases}
    \label{Eq.DefActSinglesOutQuantumConsequences}
\end{align}
A binary relation $\succeq$ over $\Acts$ satisfies behavioural state-independence if, for all $s,t\in S$ and $E,F\in\Eff$,
\begin{align}
f_{s,E}\succeq f_{s,F}\iff f_{t,E}\succeq f_{t,F}.
\label{Eq.DefBehaveStateIndependence}
\end{align}
\label{Def.BehaveStateIndependence}
\end{definition}

It is fair to read Eq.\eqref{Eq.DefBehaveStateIndependence} as saying that an agent prefers $E$ to $F$ if and only if it prefers it in all states in $S$. However, those familiar with the foundations of quantum mechanics may be tempted to offer an alternative interpretation of behavioural state-independence. By loosely identifying states of the world with possible contexts under which decisions must be made \cite{Jeffrey1965}, one may consequently see a form of non-contextuality being imposed by Def. \ref{Def.BehaveStateIndependence}. To assume this interpretation is to open the door for criticisms that we are doing a rather conservative extension of a decision-theoretical theorem. We tend to agree with this standpoint, but similar conservative extensions have been applied several times before, with surprising results \cite{LD26, FRS26, FR72, Wilce20}. We will return to this point in the next sections. For now, we need to address some technical lemmas.  


\section{Affine geometry and operator representation}\label{Sec.AffineGeometryOperatorRepresentation}

\subsection{Affine Geometry}\label{SubSec.AffineGeometry}

We start by showing a useful identity between our consequence set $\Eff$ and the set of all Hermitian matrices. Let us fix the notation first. 

\begin{definition}
    Let $X$ be a subset of a vector space $\mathbb{V}$. We define the affine hull of $X$ as follows:
    \begin{align}
        \aff(X):= \left\lbrace\sum_{i=1}^{N}a_ix_i; \,\, N \in \mathbb{N}, \,\, a_1,...,a_N \in \mathbb{R},  \right. \nonumber \\        
        \left. x_1,...,x_N \in X, \mbox{ and } \sum_{i=1}^{N}a_i=1 \right\rbrace
        \label{Eq.DefAffineHull}
    \end{align}
    \label{Def.AffineHull}
\end{definition}

\begin{lemma}[Affine hull]
    Let $\Eff$ be the consequence set given in Eq.\eqref{Def.QuantumConsequenceSpace} and $Herm(\Hilb)$ be the real vector space of all Hermitian operators over $\Hilb$, then
    \begin{align}
        \aff(\Eff)=\Herm(\Hilb).
    \end{align}
\label{lem:aff}
\end{lemma}

\begin{proof}
Because $\Eff=\{E \in Herm(\Hilb); \,\, 0 \preceq E \preceq I \}$, then for every $Z=\sum_{i=1}^{N}a_iE_i$ in $\aff(\Eff)$, with $a_1,...,a_N \in \mathbb{R}$:
\begin{align}
    Z^{\ast}=\sum_{i=1}^{N}(a_iE_i)^{\ast} = \sum_{i=1}^{N}a_iE_{i}^{\ast}=\sum_{i=1}^{N}a_iE_i = Z,
\end{align}
then, trivially, $\aff(\Eff) \subseteq \Herm(\Hilb)$. This completes the first half.

The other inclusion, $\Herm(\Hilb) \subseteq \aff(\Eff)$, is slightly less direct, as we have to show that the bounded effect interval nevertheless affinely spans the entire real vector space of Hermitian operators. 

For $Z \neq 0$ in $\Herm(\Hilb)$, choose an $a>0$ with 
    \begin{align}
        -aI\preceq Z\preceq aI.
    \end{align}
For instance, a possible choice is: 
    \begin{align}
        a:= \max \{\vert \lambda_1 \vert ,...,\vert \lambda_{N} \vert \},
    \end{align}
where $\lambda_1,...,\lambda_{N}$ are the eigenvalues of $Z$. Now, define two auxiliary operators $E_{\pm}$ as:
    \begin{align}
        E_+=\frac{Z+aI}{2a},\qquad E_- =\frac{-Z+aI}{2a}.
    \end{align}
Then $E_\pm\in\Eff$ and
    \begin{align}
        Z=aE_+-aE_-+1\cdot0,
        \label{Eq.AffineCombinationForEPM}
    \end{align}
The coefficients in the sum above are $a$, $-a$, and $1$, and because $a-a+1=1$, Eq.\eqref{Eq.AffineCombinationForEPM} is an affine combination of effects. Therefore, $Z \in \aff(\Eff)$.
\end{proof}

There is a useful consequence from the previous lemma that should be emphasised. Recall that $S$ is a finite set of states of the world, with cardinality $\vert S \vert = M$, so that we can readily identify an act $f \in \Acts$ with a  vector $\vec{f} \in \Eff^{M}$ whose components are:
    \begin{align}
        \vec{f} = \left( f(s_1),f(s_2),...,f(s_M) \right) \in \Eff^{M}.
        \label{Eq.DefVectorOfActs}
    \end{align}
Consequently, affine combinations of acts in $\Acts$ are equivalently identified with vectors whose components are affine combinations of effects in $\Eff$ and, therefore, with vectors of Hermitian matrices. That is:
\begin{align}
    f \in \aff(\Acts) & \iff f = \sum_{i=1}^{N}a_i f_i \nonumber  \\
    & \iff \vec{f} = \left(\sum_{i=1}^{N}a_i f_i(s_1),...,\sum_{i=1}^{N}a_i f_i(s_M)  \right) \nonumber \\
    & \iff \vec{f} \in \aff(\Eff)^{M} \nonumber \\
    & \iff \vec{f} \in \Herm(\Hilb)^{M}
    \label{Eq.AffineCombinationsOfActsAffineCombinationofEffs}
\end{align}
Finally, if we identify vectors in $\Herm(\Hilb)^{M}$ with functions in $\Herm(\Hilb)^{S} = \{f:S \to \Herm(\Hilb)\}$, like we did before, we can promptly write:
\begin{align}
    \aff(\Acts) \simeq \Herm(\Hilb)^S.
\end{align}
We organise this discussion in a corollary:
\begin{corollary}
Let $S$ be a finite set of states of the world. The following equality holds true
    \begin{align}
     \aff(\Acts)=\Herm(\Hilb)^S,
    \label{Eq.CoroActsHerm}
\end{align}
up to an isomorphism.
\label{Coro.ActsHerm}
\end{corollary}


The previous corollary is fundamental for determining the form of each affine utility function \(U_\alpha:\Acts\to\mathbb{R}\) occurring in Theorem~\ref{Thm.MultiValuedMixtureRep}. We now make explicit how the affine structure of the act space allows us to extend each such utility to the entire real vector space \(\Herm(\Hilb)^S\).

\begin{proposition}
    Let $\{U_{\alpha}\}_{\alpha \in L}$ be a family of affine functions defined on $\Acts$ as in Thm.\ref{Thm.MultiValuedMixtureRep}, then each $U_{\alpha}$ can be uniquely extended to an affine function $U_{ext}$ whose domain is $\Herm(\Hilb)^{S}$. 
    \label{Prop.ExtendingUAlphas}
\end{proposition}
\begin{proof}
Fix an index \(\alpha\in L\), and write \(U=U_\alpha\). By Theorem~\ref{Thm.MultiValuedMixtureRep}, \(U\) is affine on \(\Acts\). In particular, it is also convex and for every \(f,g\in\Acts\) and every \(\lambda\in[0,1]\),
    \begin{align}
        U(\lambda f+(1-\lambda)g)
        =\lambda U(f)+(1-\lambda)U(g).
    \end{align}
We stress that, at this stage, \(U\) is defined only on the original act space \(\Acts\), not on its affine hull. Let
    \begin{align}
        \mathbb{V}:=\Herm(\Hilb)^S.
    \end{align}
By Corollary~\ref{Coro.ActsHerm}, the affine hull of $\Acts$ is $\mathbb{V}$. Thus, every $f\in \mathbb{V}$ can be expressed as a finite affine combination of acts: there exist $f_1,\ldots,f_n \in \Acts$ and $a_1,\ldots,a_n \in \mathbb{R}$ such that
\begin{align}
f=\sum_{i=1}^n a_i f_i, \mbox{ and }
\sum_{i=1}^n a_i=1.
\end{align}
This suggests defining an extension of \(U\) to $\mathbb{V}$ by setting
\begin{align}
U_{ext}(f):=\sum_{i=1}^n a_i U(f_i).
\label{Eq.TentativeExtension}
\end{align}
There is, however, a subtle point that must be addressed before we settle on this definition: an element of an affine hull need not have a unique representation as an affine combination. We must therefore establish that the value assigned by Eq.\eqref{Eq.TentativeExtension} is independent of the particular affine representation chosen.

Suppose that the same $f \in \mathbb{V}$ admits two affine representations, say,
\begin{align}
\sum_{i=1}^n a_i f_i = f 
=\sum_{j=1}^m b_j g_j, \mbox{ where }
\sum_{i=1}^n a_i=1=\sum_{j=1}^m b_j,
\end{align}
with all $f_i,g_j$ belonging to $\Acts$. Subtracting these two expressions gives
\begin{align}
\sum_{i=1}^n a_i f_i-\sum_{j=1}^m b_j g_j=0,
\mbox{ and }
\sum_{i=1}^n a_i-\sum_{j=1}^m b_j=0.
\end{align}
For convenience, organise the acts in the two representations into a single finite list \(h_1,\ldots,h_{n+m}\), and let \(c_1,\ldots,c_{n+m}\) denote their corresponding signed coefficients. We then have
\begin{align}
\sum_{k=1}^{n+m}c_k h_k=0,
\mbox{ and }
\sum_{k=1}^{n+m}c_k=0.
\label{Eq.List}
\end{align}
Now, write $c_k^+=\max\{c_k,0\}$ and $c_k^-=\max\{-c_k,0\}$. The second equality in Eq.\eqref{Eq.List} implies
\begin{align}
r:=\sum_{k=1}^{n+m}c_k^+
=\sum_{k=1}^{n+m}c_k^-.
\end{align}
If $r=0$, then every $c_k^+$ and $c_k^-$ is zero, so every $c_k=0$ and the desired equality of utility values follows immediately. We may therefore assume that $r>0$. The first equality in Eq.\eqref{Eq.List} then guarantees that:
\begin{align}
\sum_{k=1}^{n+m}\frac{c_k^+}{r}h_k
=
\sum_{k=1}^{n+m}\frac{c_k^-}{r}h_k.
\label{Eq.AuxiliaryEquationCwithH}
\end{align}
Both sides of Eq.\eqref{Eq.AuxiliaryEquationCwithH} are convex combinations of acts. Since $\Acts$ is convex, they are themselves acts, and the affinity of \(U\) therefore gives
\begin{align}
\sum_{k=1}^{n+m}\frac{c_k^+}{r}U(h_k)
=
\sum_{k=1}^{n+m}\frac{c_k^-}{r}U(h_k).
\end{align}
Multiplying by \(r\) and rearranging, we obtain
\begin{align}
\sum_{k=1}^{n+m}c_k U(h_k)=0.
\end{align}
Consequently,
\begin{align}
\sum_{i=1}^n a_i U(f_i)
=
\sum_{j=1}^m b_j U(g_j).
\end{align}
This proves that the tentative definition in Eq.\eqref{Eq.TentativeExtension} is really independent of the chosen affine representation. Hence $U_{ext}:V\to\mathbb{R}$ is well-defined.

Moreover, we affirm that $U_{ext}$ is affine. Indeed, let $f,g \in \mathbb{V}$, and choose affine representations
    \begin{align}
        f=\sum_{i=1}^n a_i f_i,
        \mbox{ and }
        g=\sum_{j=1}^m b_j g_j.
\end{align}
For any $\lambda\in[0,1]$, we have
\begin{align}
\lambda f+(1-\lambda)g
=
\sum_{i=1}^n\lambda a_i f_i
+
\sum_{j=1}^m(1-\lambda)b_j g_j.
\end{align}
The coefficients on the right-hand side sum to one. By the definition of the extension and its well-definedness,
\begin{align}
U_{ext}&(\lambda f+(1-\lambda)g) = \nonumber \\
&=\lambda\sum_{i=1}^n a_iU(f_i)
 +(1-\lambda)\sum_{j=1}^m b_jU(g_j) \nonumber \\
&=\lambda U_{ext}(f)+(1-\lambda) U_{ext}(g).
\end{align}
Thus $U_{ext}$ is affine on the whole vector space $\mathbb{V}$. 

Note that, as expected, this extension also agrees with the original utility on \(\Acts\). For any \(f\in\Acts\), the representation consisting of the single act \(f\), with coefficient one, gives $U_{ext}(f) = U(f)$.

Finally, we want to show that the extension is unique as well. To see this, suppose that $\widetilde U:V\to\mathbb{R}$ is another affine function agreeing with $U$ on $\Acts$. For any $f \in \mathbb{V}$, choose an affine representation $f=\sum_{i}^{n} a_i f_i$. Affinity of \(\widetilde U\) implies
\begin{align}
\widetilde U(f)=\sum_{i=1}^{n} a_i\widetilde U(f_i)
=\sum_i a_i U(f_i)
=U_{ext}(f).
\end{align}
Since this holds for every $f \in \mathbb{V}$, we conclude that
    \begin{align}
        \widetilde U=U_{ext}.
\end{align}
\end{proof}

We have therefore established that every affine supporting utility on \(\Acts\) admits a unique affine extension to the whole space. In this way, every utility appearing in Thm.\ref {Thm.MultiValuedMixtureRep}, whose initial domain is $\Acts$, can be adequately extended to $\Herm(\Hilb)^S$. The question now is: how about their form?

\subsection{Operator Representation}\label{SubSec.OperatorRepresentation}

We can now use the finite-dimensional structure of $\mathbb{V}:=\Herm(\Hilb)^S$ to obtain the desired operator representation for the utilities in Thm.\ref{Thm.MultiValuedMixtureRep}. To begin with, we define the special act that assigns the zero effect to every state:

\begin{definition}[Null Act]
    The null act in $\Acts$, denoted by $f_{0}$ is defined as follows:
    \begin{align}
        f_{0}: S &\to \Eff, \nonumber \\
            s &\mapsto E_{0}
        \label{Eq.DefNullAct}
    \end{align}
    where $E_{0}$ is the null effect, $E_{0}(\psi)=0$ for every $\psi$ in $\Hilb$.
\end{definition}

Since $f_{0} \in \Acts \subseteq \aff(\Acts)$, it follows that $f_{0} \in \mathbb{V}$. Define, then, 
\begin{align}
L(f):= U_{ext}(f)- U_{ext}(f_{0}),
\qquad f\in \mathbb{V}.
\end{align}
Because $U_{ext}$ is affine and $\mathbb{V}$ is a vector space, $L$ is linear. Now, if we set $c_U:=U_{ext}(f_{0})$, we consequently have
\begin{align}
U_{ext}(f) = c_U + L(f),
\qquad f \in \mathbb{V}.
\label{Eq.FinalFormForUwithCuandL}
\end{align}

We are almost done. To represent this linear functional, equip $\mathbb{V}$ with a real inner product.
    \begin{definition} [Pairing Product]
    For every $f,g$ in $\mathbb{V}=\Herm(\Hilb)^{S}$, define its pairing product $\langle f,g\rangle$ as:
        \begin{align}
            \langle f,g\rangle
            :=
            \sum_{s\in S}\Tr(f(s)g(s)),
            \qquad f,g \in  \mathbb{V}.
            \label{Eq.DefInnerProduct}
        \end{align}
        \label{Def.InnerProduct}
\end{definition}
This is a valid inner product because the usual Hilbert--Schmidt pairing $\langle E, F \rangle = \Tr(E^{\ast}F)$ is real-valued and positive definite on the real vector space of Hermitian operators. 

Finally, since $\mathbb{V}$ is finite-dimensional, the Riesz representation theorem \cite{HK71} guarantees that there is a unique $B^{U} \in \mathbb{V}$ such that
\begin{align}
\forall f:    L(f)=\langle B^U,f\rangle
=\sum_{s\in S}\Tr(B_s^U f(s)),
\end{align}
where \(B_s^U\in\Herm(\Hilb)\) denotes the component of \(B^U\) corresponding to state $s$---recall definition \ref{Def.InnerProduct}. Combining this identity with Eq.\eqref{Eq.FinalFormForUwithCuandL}, and restricting back to the original act space, when necessary, we finally obtain
\begin{align}
U(f)=c_U+\sum_{s\in S}\Tr(B_s^U f(s)),
\mbox{ for every } f \in \Acts.
\label{Eq.FormOfUInvolvingBUs}
\end{align}

We summarise this discussion with the following result

\begin{theorem}[Multi-Utility Representation Theorem - Second Version]
    If $\succeq$ is a suitable preference over $\Acts$, then there exists an index set $L$, with $|L| \geq 1$, a family of real-valued constants $\{c_{\alpha}\}_{\alpha \in L}$, and a doubly-indexed family of Hermitian operators $\{B_{s}^{\alpha}\}_{(s,\alpha) \in S \times L}$ in $\Herm(\Hilb)$  such that 
    \begin{align}
        f \succeq g \iff \forall \alpha \in L: \,\, U_{\alpha}(f) \geq U_{\alpha}(g),
        \label{Eq.ThmMultiValuedMixtureRep2ndVersion}
    \end{align}
    where,
    \begin{align}
        U_{\alpha}(f) = c_{\alpha}+\sum_{s\in S}\Tr(B_{s}^{\alpha} f(s))
    \end{align}
\label{Thm.MultiValuedMixtureRep2ndVersion}    
\end{theorem} 

One way to read Thm.\ref{Thm.MultiValuedMixtureRep2ndVersion} is that every affine utility admits a statewise representation by Hermitian coefficient operators. Notice that this operator representation follows from $(i)$ the affine structure of the act space and $(ii)$ the finite-dimensionality of $\Hilb$---more specifically from the Riesz theorem. Also note that no positivity or proportionality property of the operators $B_{s}^{\alpha}$ has yet been established (the second half of our main question); those properties will require additional assumptions introduced in the following sections.

\section{Statewise proportionality and factorisation}\label{Sec.ProportionalityFactorisation}

The previous section explains where the operators $B_s^{\alpha}$ come from. The next step is to investigate what makes them separate into a classical-like component and a common quantum-like component. This separation is a strong feature, and to obtain it we impose a sort of state-independence at the level of each utility functional---an operational restriction on the utilities rather than on the preference itself, as we had done with defs.\ref{Def.IncompletePreference}-\ref{Def.MixtureContinuity}. This is the precise mathematical bridge between the behavioural idea introduced above and the desired Anscombe--Aumann form.

\begin{definition}[State-independence for affine utilities]
Let $\mathcal{U}=\{U_{\alpha}\}_{\alpha \in L}$ be a family of utility functionals whose affine representation is given by:
\begin{align}
U_{\alpha}(f)=c_{\alpha}+\sum_{s\in S}\Tr(B_s^{\alpha} f(s)).
\label{Eq.FunctionalFormWithBs}
\end{align}
We say that $\mathcal{U}$ respects \emph{state-independence at the utility level}, when for each $\alpha \in L$, and for every $s,t\in S$ and $E,F\in\Eff$,
\begin{align}
\Tr(B_s^{\alpha}(E-F)) \geq 0 \iff \Tr(B_t^{\alpha}(E-F)) \geq 0,
\end{align}
and that the corresponding functionals are nonzero.
\label{Def.StateIndependenceUtilityLevel}
\end{definition}

\begin{proposition}
The utility-level condition implies behavioural state-independence for a suitably represented preference.
\label{Prop.UtilityIndImpliesBehaviourInd}
\end{proposition}
\begin{proof}
    We present a sketch of the proof
\begin{align}
f_{s,E}\succeq f_{s,F}
&\iff
\Tr(B_s^\alpha(E-F))\geq 0
\ \forall \alpha\in L
\nonumber\\
&\iff
\Tr(B_t^\alpha(E-F))\geq 0
\ \forall \alpha\in L
\nonumber\\
&\iff
f_{t,E}\succeq f_{t,F}.
\end{align}
   The first and last equivalence come from the fact that we can represent the suitable preference relation by eq.\eqref{Eq.FunctionalFormWithBs} whereas the middle equivalence is precisely the assumed state-independence for affine utilities.
\end{proof}

Note that we do not establish the converse of proposition \ref{Prop.UtilityIndImpliesBehaviourInd}. Without additional assumptions, there is no guarantee that behavioural state-independence carries over to the utility-level condition. This is precisely why we are at pains to separate those two hypotheses in our construction.

The operational definition above is essential to recover the intended Anscombe--Aumann interpretation of state-independence. We will assume this stronger form of state independence because it is key to our proof—the factorisation strategy we came up with, unfortunately, only works on a utility-by-utility basis.
What should be emphasised, though, is that under completeness, the distinction between state-independence at the preference level and state-independence at the utility level is largely hidden; under incompleteness, we had to make it more explicit.

\begin{lemma}[Positive Proportionality]
Under the state-independence condition of Def.~\ref{Def.StateIndependenceUtilityLevel}, for a fixed $\alpha \in L$ and every pair of states $s,t\in S$, there exists a unique $\lambda_{st}^{\alpha}>0$ such that
\begin{align}
B_s^{\alpha}=\lambda_{st}^{\alpha} B_t^{\alpha}.
\label{Eq.LemmaPositiveProportionality}
\end{align}
\label{LemmaPositiveProportionality}
\end{lemma}
\begin{proof}
Fix $\alpha \in L$ and two states $s,t\in S$. Define the linear functionals
\begin{align}
L_s(Z):=\Tr(B_s^{\alpha} Z)
\mbox{ and }
L_t(Z):=\Tr(B_t^{\alpha} Z),
\end{align}
for every $Z \in \Herm(\Hilb)$. By the state-independence condition, for every pair of effects $E,F\in\Eff$,
\begin{align}
L_s(E-F)\geq0
\iff
L_t(E-F)\geq0.
\label{Eq.CommonSignOrdering}
\end{align}

We first extend this common sign ordering to arbitrary Hermitian operators. To do so, let $Z \in \Herm(\Hilb)$ be an arbitrary Hermitian matrix. Since $Z$ is Hermitian, its eigenvalues are real. Hence, for sufficiently small $\varepsilon>0$, the operator
\begin{equation}
E:=\frac{1}{2} I+\varepsilon Z
\end{equation}
is a valid effect.  Now, by setting $F=\frac{1}{2} I$  in \eqref{Eq.CommonSignOrdering}, we obtain
\begin{align}
L_s(\varepsilon Z)\geq0
\iff
L_t(\varepsilon Z)\geq0.
\end{align}
Since $\varepsilon>0$ and both functionals are linear, it follows that
\begin{align}
L_s(Z)\geq0
\iff
L_t(Z)\geq0,
\label{Eq.ExtensionForHermitianMatrices}
\end{align}
for every $Z \in \Herm(\Hilb)$.

We next show that the two functionals have the same kernel. Suppose that $Z \in \ker (L_t)$, so that $L_t(Z)=0$. By \eqref{Eq.ExtensionForHermitianMatrices}, $L_s(Z)\geq0$. On the other hand, by applying the same reasoning to $-Z$ we also obtain $L_s(-Z) \geq 0$. Therefore $L_s(Z)=0$, and hence $\ker (L_t) \subseteq \ker(L_s)$. Swapping $s$ and $t$ gives the reverse inclusion. Thus,
\begin{align}
\ker(L_s) = \ker(L_t).
\label{Eq.TheKernelsAreTheSame}
\end{align}

Because $L_t$ is nonzero, we may therefore choose an $X_0\in\Herm(\Hilb)$ such that $L_t(X_0)>0$. By \eqref{Eq.ExtensionForHermitianMatrices}, it also holds true that $L_s(X_0)\geq0$. In fact, this inequality is strict: otherwise $ X_0 \in \ker(L_s)=\ker(L_t)$, contradicting $L_t(X_0)>0$. Consequently, the ratio
\begin{equation}
\lambda_{st}^{\alpha}:= \frac{L_s(X_0)}{L_t(X_0)}
\end{equation}
is well-defined and strictly positive—recall that each $U_{\alpha}$ gives rise to such family of linear functionals $L_{s}$. Now, let $Z \in \Herm(\Hilb)$ be arbitrary, and set
\begin{align}
Y:=Z-\frac{L_t(Z)}{L_t(X_0)}X_0.
\end{align}
By linearity,
\begin{align}
L_t(Y) =L_t(Z)-\frac{L_t(Z)}{L_t(X_0)}L_t(X_0)
=0.
\end{align}
Thus $Y \in \ker(L_t)=\ker(L_s)$, and therefore $L_s(Y)=0$. Expanding this last equality:
\begin{align}
L_s(Z) = \frac{L_s(X_0)}{L_t(X_0)}L_t(Z) = \lambda_{st}^{\alpha} L_t(Z).
\end{align}
Consequently, for every $X \in \Herm(\Hilb)$
\begin{align}
\Tr\left[
\left(B_s^{{\alpha}}-\lambda_{st}^{\alpha} B_t^{\alpha}\right)X
\right]=0.
\end{align}
Therefore,
\begin{align}
B_s^{\alpha}=\lambda_{st}^{\alpha} B_t^{\alpha},
\mbox{ with } \lambda_{st}^{\alpha}>0.
\end{align}

Finally, the uniqueness part follows from the fact that $B_t^{\alpha}\neq0$. Indeed, if another $\mu_{st}^U>0$ satisfied
$B_s^{\alpha}=\mu_{st}^{\alpha} B_t^{\alpha}$, then
\begin{align}
(\lambda_{st}^{\alpha}-\mu_{st}^{\alpha})B_t^{\alpha}=0,
\end{align}
which forces $\lambda_{st}^{\alpha}=\mu_{st}^{\alpha}$.
\end{proof}

Now it should be clear why such a representation-level condition is not a direct consequence of reflexivity, transitivity, strong independence, mixture-continuity, and Loewner monotonicity alone. For example, with two states and $\Hilb=\mathbb C^2$, take $B_1=I$ and $B_2=\operatorname{diag}(1,0)$. The resulting affine preference is complete, strongly independent, mixture-continuous, and Loewner-monotone (more on that in a minute), but the two coefficient operators are not proportional. Hence a separate state-independence restriction is genuinely needed for the factorisation argument.

\begin{proposition}[Classical--quantum factorisation]
Fix $\alpha \in L$. If all $B_s^{\alpha}$ are nonzero and satisfy positive proportionality, then there exist $p_{\alpha} \in \Delta(S)$ and a nonzero Hermitian operator $\widetilde B_{\alpha} \in \Herm(\Hilb)$ such that
    \begin{align}
        B_s^{\alpha}=p_{\alpha}(s)\widetilde B_{\alpha}.
        \label{Prop.ClassicalQuantumFactorisation}
    \end{align}
\end{proposition}
\begin{proof}
Choose an $s_0 \in S$. By the previous lemma, we know that for every $s \in S$, there exists an $r(s)>0$ such that $B_s^{\alpha}=r(s)B_{s_{0}}^{\alpha}$. Now, define $R:=\sum_{s \in S} r(s)$, $p_\alpha(s):=r(s)/R$, and $\widetilde{B}_{\alpha}=RB_{s_{0}}^{\alpha}$. Then $p_{\alpha} \in \Delta(S)$ and $B_s^{\alpha}=p_{\alpha}(s)\widetilde{B_{\alpha}}$.
\end{proof}

\section{Positivity and density normalisation}\label{Sec.PositivityAndDensityNormalisation}

\subsection{Positivity under Monotonicity}

The factorisation seen in Eq.\eqref{Prop.ClassicalQuantumFactorisation} identifies the two roles we wanted to separate. However, while $p_{\alpha}$ belongs to $\Delta(S)$ as expected, its operator counterpart, the $\widetilde B_{\alpha}$ part, is not guaranteed to be a quantum state. We need, then, to impose yet another layer on our framework. We do so by demanding a specific kind of order-behavioural input: Loewner monotonicity—it acts so as to guarantee that a larger effect must never be assigned a smaller value. In finite dimensions, this forces the common operator to be positive, after which trace normalisation turns it into a density operator.

\begin{definition}[Loewner Monotonicity]
We say that $\succeq$ satisfies \emph{Loewner monotonicity} whenever $F - E$ is positive semidefinite implies that the constant act with consequence $F$ is preferred to the constant act with consequence $E$. 
\end{definition}

\textbf{Remark:} A constant act with consequence $E$ in $\Eff$ is defined as
    \begin{align}
        C_{E}: S &\to \Eff, \nonumber \\
            s &\mapsto E.
        \label{Eq.DefConstantAct}
    \end{align}
If we identify $\Eff^{S}$ with $\Eff^{|S|}$, then the constant act can be simply written as 
\begin{align}
    C_{E}=\underbrace{(E,E,...,E)}_{|S|-times}.
\end{align}
In this sense, if we write $F \succeq E$ whenever $F - E$ is positive semidefinite, we can rewrite the Loewner monotonicity definition as follows:
\begin{align}
    F \succeq E \mbox{ as effects } \Rightarrow C_{F} \succeq C_{E} \mbox{ as acts}.
\end{align}

\begin{proposition}[Positivity]
Under Loewner monotonicity, every $\widetilde{B}_{\alpha}$ is semidefinite positive.
\end{proposition}
\begin{proof}
The proof goes by contradiction. Suppose that this were not the case, that is, suppose that $\widetilde{B}_{\alpha}$ were not semidefinite positive. It would be possible, then, to choose a unit vector $\ket\psi$ with $\langle\psi|\widetilde B_{\alpha}|\psi\rangle<0$. The projector $P_\psi=\ket\psi\bra\psi$ satisfies $P_\psi \succeq_{\mathrm L} 0$. Monotonicity would require the constant act $P_\psi$ to be (weakly) preferred to $0$, but
\begin{align}
U_\alpha(C_{P_\psi})-U_\alpha(C_0)=\Tr(\widetilde B_{\alpha} P_\psi)=\langle\psi|\widetilde B_{\alpha}|\psi\rangle<0,
\end{align}
a contradiction.
\end{proof}

The proposition above is essentially what we needed, but before going ahead, we can also use it as part of the argument to distinguish Loewner monotonicity from utility state-independence. While the former determines the sign of each supporting quantum functional, the latter separates that common quantum functional from the classical state weights. Therefore, in general, these two properties have logically distinct consequences.  

\subsection{Positivity}

We have proven that under Loewner monotonicity, each $\widetilde{B}_{\alpha}$ is semidefinite positive. To conclude this part of the argument, it remains to show that they can also be appropriately normalised. This is fairly simple. To do so, define an auxiliary $\rho_{\alpha}$ as: 
\begin{align}
    \rho_{\alpha}=\frac{\widetilde B_{\alpha}}{\Tr(\widetilde B_{\alpha})}.
\end{align}
Because $\widetilde B_{\alpha}$ is positive semidefinite and nonzero, $\Tr(\widetilde B_{\alpha})>0$. Hence $\rho_{\alpha}\succeq0$ and $\Tr(\rho_{\alpha})=1$, so $\rho_{\alpha}$ is a density operator. 

Finally, recall that we are always interested in comparisons like $U_{\alpha}(f) \geq U_{\alpha}(g)$, where each $U_{\alpha}$ is supposed to be of the form given in Eq.\eqref{Eq.FunctionalFormWithBs} with the factorisation of Eq.\eqref{Prop.ClassicalQuantumFactorisation}, so that any positive rescaling does not change the ordering. Consequently, replacing each affine utility by its normalised representative does not change the represented preference, and we can focus on the following family of utility functionals:
    \begin{align}
        U_{p_{\alpha},\rho_{\alpha}}(f)=\sum_s p_{\alpha}(s)\Tr(\rho_{\alpha} f(s)).
    \label{Eq.FunctionalFormWithPsandQsSeparated}
    \end{align}

We summarise our findings so far in the theorem and corollary below:

\begin{theorem}[Quantum--classical affine representation]
Suppose that $\succeq$ is a suitable preference and let $\{U_\alpha\}_{\alpha\in L}$ be a nonempty affine multi-representing family supplied by Theorem~\ref{Thm.MultiValuedMixtureRep}. Suppose, in addition, that this chosen family satisfies the representation-level state-independence condition of Definition~\ref{Def.StateIndependenceUtilityLevel}, and that $\succeq$ satisfies Loewner monotonicity. Then, for every $\alpha\in L$, there exist $p_\alpha\in\Delta(S)$, $\rho_\alpha\in\Dens(\Hilb)$, $a_\alpha>0$, and $b_\alpha\in\mathbb R$ such that
\begin{align}
U_\alpha(f)&=b_\alpha+a_\alpha U_{p_\alpha,\rho_\alpha}(f), \mbox{ with }\\ \nonumber \\
U_{p_\alpha,\rho_\alpha}(f)&=\sum_{s\in S}p_\alpha(s)\Tr(\rho_\alpha f(s)).
\label{Eq.FunctionalFormWithPsandQsSeparatedv2}
\end{align}
In particular, $U_\alpha$ is positively affinely equivalent to the normalised utility $U_{p_\alpha,\rho_\alpha}$.
\label{Thm.FunctionalFormWithPsandQsSeparatedv2}
\end{theorem}

\begin{proof}
Fix $\alpha\in L$. By the statewise operator representation,
\begin{align}
U_\alpha(f)=c_\alpha+\sum_{s\in S}\Tr(B_s^\alpha f(s)).
\end{align}
The representation-level state-independence assumption and the positive-proportionality lemma imply that the nonzero operators $B_s^\alpha$ are positive scalar multiples of one another. The factorisation proposition therefore gives $p_\alpha\in\Delta(S)$ and a nonzero Hermitian operator $\widetilde B_\alpha$ such that $B_s^\alpha=p_\alpha(s)\widetilde B_\alpha$ for every $s$. Loewner monotonicity implies $\widetilde B_\alpha\succeq0$. Since it is nonzero and positive semidefinite, its trace is strictly positive. Define
\begin{align}
\rho_\alpha:=\frac{\widetilde B_\alpha}{\Tr(\widetilde B_\alpha)}.
\end{align}
Then $\rho_\alpha$ is a density operator, and
\begin{align}
U_\alpha(f)=c_\alpha+\Tr(\widetilde B_\alpha)U_{p_\alpha,\rho_\alpha}(f).
\end{align}
Thus the required constants are $b_\alpha=c_\alpha$ and $a_\alpha=\Tr(\widetilde B_\alpha)>0$.
\end{proof}

Moreover, the nonzero functional clause in representation-level state-independence ensures that every proportionality factor is strictly positive. Hence each probability distribution $p_\alpha$ obtained directly from the chosen supporting family has full support. This full-support conclusion need not hold for every additional pair in the canonical maximal compatible family introduced below.

\begin{corollary}
Under the conditions of Thm.\ref{Thm.FunctionalFormWithPsandQsSeparatedv2}, there is a nonempty family
\begin{align}
\mathcal{R} \subseteq \Delta(S) \times \Dens(\Hilb)
\end{align}
such that
\begin{align}
f\succeq g\iff U_{p,\rho}(f)\ge U_{p,\rho}(g), \,\, \forall \,\,(p,\rho) \in \mathcal{R}.
\label{Eq.CoroAdmissibleFamily}
\end{align}
    \label{Coro.AdmissibleFamily}
\end{corollary}
\begin{proof}
Apply the theorem to every member of the nonempty representing family. Positive affine transformations preserve each utility's ordering, so the resulting nonempty family of normalised pairs represents the same preference.
\end{proof}

Together, theorem \ref{Thm.FunctionalFormWithPsandQsSeparatedv2} and corollary \ref{Coro.AdmissibleFamily} answer the two main questions we set out to address: under the stated hypotheses, we can now represent a preference on $\Acts$ via a family of utility functionals that splits into a classical-like contribution and a quantum-like part—granting both $p$'s and $\rho$'s a subjective character—on par with Anscombe and Aumann's original model \cite{AA63}. However, we can also address side-questions within this formalism. The fact that we have now opted to rewrite the family of affine functionals $\{U_{\alpha}\}_{\alpha \in L}$ as $\{U_{p,\rho}\}_{(p,\rho) \in \mathcal{R}}$ to make explicit the dependence of $(p,\rho)$ in an admissible set $\mathcal{R}$ is not just a matter of aesthetics, as we will see in the next section.

\section{Admissibility and Slices of Acts}\label{Sec.AdmissibilityAndSlices}

The preceding argument constructs normalised pairs from a chosen supporting family. It is useful to distinguish that construction from the question of which pairs are compatible with the preference. Incomplete preferences generally admit many supporting pairs, some of which may be redundant \cite{Fritz17}. In this section, to investigate this side question, we move from the existence and the form of an ordering representation to what the utility representation implies for a valid factorised pair $(p,\rho) \in  \Delta(S)\times\Dens(\Hilb)$.

\subsection{Admissibility}\label{SubSec.Admissibility}

\begin{definition}[Representation family]
A nonempty $\mathcal R\subseteq\Delta(S)\times\Dens(\Hilb)$ is a \emph{representation family} whenever the following holds:
\begin{align}
f\succeq g\iff U_{p,\rho}(f)\ge U_{p,\rho}(g), \,\, \forall \,\,(p,\rho) \in \mathcal{R}.
\label{Eq.DefCoroAdmissibleFamily}
\end{align}
We say that a pair $(p,\rho)$ in $\Delta(S) \times \Dens(\Hilb)$ is \emph{admissible} when it belongs to a representation family $\mathcal{R}$.
\label{Def.AdmissibleFamily}
\end{definition}

\begin{definition}[Canonical maximal compatible family]
Let $\succeq$ be a preorder over $\Acts$. Define the set of all normalised pairs that respect every comparison in $\succeq$ by
\begin{align}
\Rmax=\{(p,\rho) \in & \Delta(S) \times \Dens(\Hilb):
U_{p,\rho}(f)\ge U_{p,\rho}(g) &  \nonumber \\
 &\text{ whenever }f\succeq g,\text{ for all }f,g\in\Acts\}.
\label{Eq.DefCanonicalMaximalCompatibleFamily}
\end{align}
\label{Def.CanonicalMaximalCompatibleFamily}
\end{definition}

\begin{proposition}[Maximal compatible family]
If $\mathcal{R}_0$ represents $\succeq$, then $\mathcal{R}_0 \subseteq \Rmax$. Also, $\Rmax$ itself represents $\succeq$.
\label{Prop.MaximalCompaitbleFamily}
\end{proposition}
\begin{proof}
First, every pair in a representing family respects every comparison, so $\mathcal{R}_0\subseteq\Rmax$. Now we need to prove that $\Rmax$ is also a representing family. If $f\succeq g$, by definition, all pairs in $\Rmax$ respect the comparison. On the other hand, by contradiction, if $U_{p,\rho}(f)\geq U_{p,\rho}(g), \,\, \forall \,\,(p,\rho) \in \Rmax$ and $f \not\succeq g$, there would exist a pair $(p_{0},\rho_{0}) \in \mathcal{R}_{0}$ such that $U_{p_{0},\rho_{0}}(f) < U_{p_{0},\rho_{0}}(g)$; a contradiction because $\mathcal{R}_{0} \subseteq \Rmax$.
\end{proof}

From this point onward, unless otherwise stated, we take $\mathcal{R}=\Rmax$. This will permit to focus on general properties of admissibility rather than on particular features arising from selecting only a subset of representing pairs. Consequently, however, the maximal family may contain additional compatible pairs beyond those constructed directly from the supporting utilities. Since the original representing family is contained in $\Rmax$, the latter contains at least one full-support probability distribution under the hypotheses of the factorisation theorem. Additional pairs in $\Rmax$ need not have full support themselves. Recall that properties established for every pair in $\Rmax$ also hold for every subfamily of $\Rmax$.

The next lemma shows that the utility representation uniquely determines the representation pairs.  

\begin{lemma}[Identifiability of an admissible pair]
If two admissible pairs $(p,\rho)$ and $(p',\rho')$ in $\Delta(S)\times\Dens(\Hilb)$ induce the same (weak) ordering on all acts, then $p=p'$ and $\rho=\rho'$.
\end{lemma}

\begin{proof}
Let
\begin{align}
L(Z):=\sum_{s\in S}p(s)\Tr(\rho Z_s),
\mbox{ and }
L'(Z):=\sum_{s\in S}p'(s)\Tr(\rho'Z_s),
\end{align}
be two linear functionals defined for every $Z=(Z_s)_{s\in S}$ in $\Herm(\Hilb)^{|S|} \simeq \aff(\Acts) \supseteq \Acts$ (see corollary \ref{Coro.ActsHerm}). By hypothesis, we assume that the two normalised pairs induce the same (weak) ordering on all acts. Then, for every pair of acts $f,g\in\Acts$,
\begin{align}
L(f-g)\geq0
\iff
L'(f-g)\geq0.
\label{Eq.FunctionalLOverActs}
\end{align}

We first extend this common sign ordering to the entire vector space $\Herm(\Hilb)^S$. To do so, consider the interior act
\begin{align}
f_{\mathrm{int}}:=\left(\frac12 I,\ldots,\frac12 I\right).
\end{align}
Similarly to what we did before, in lemma \ref{lem:aff}, for an arbitrary $Z=(Z_s)_{s\in S}\in\Herm(\Hilb)^{|S|}$, finiteness of the set of the states of the world $S$ and boundedness of the eigenvalues of each Hermitian operator $Z_s$ guarantee that, for sufficiently small $\varepsilon>0$,
\begin{align}
f_\varepsilon:=f_{\mathrm{int}}+\varepsilon Z
\end{align}
is an act. Setting $f=f_\varepsilon$ and $g=f_{\mathrm{int}}$ in \eqref{Eq.FunctionalLOverActs}, we obtain
\begin{align}
L(\varepsilon Z)\geq0
\iff
L'(\varepsilon Z)\geq0.
\end{align}
By linearity and $\varepsilon>0$, this is equivalent to
\begin{align}
L(Z)\geq0
\iff
L'(Z)\geq0.
\label{Eq.ExtenstionForL}
\end{align}

for every $Z \in\Herm(\Hilb)^S$. Now, applying \eqref{Eq.ExtenstionForL} to both $Z$ and $-Z$ shows that
\begin{align}
\ker(L)=\ker(L').
\end{align}
Moreover, both functionals are nonzero, since
\begin{align}
L(I,\ldots,I)
=
\sum_{s\in S}p(s)\Tr(\rho I)=1
\end{align}
and, analogously, $L'(I,\ldots,I)=1$. It follows from the common-kernel argument (see lemma \ref{LemmaPositiveProportionality}) that there exists $\lambda>0$ such that
\begin{align}
L(Z)=\lambda L'(Z)
\label{Eq.ProportionalityForLs}
\end{align}
for every $Z \in \Herm(\Hilb)^{|S|}$. Evaluating Eq.\eqref{Eq.ProportionalityForLs} for $X=(I,\ldots,I)$ and using the normalisation of $p,p' \in \Delta(S)$ and of $\rho,\rho' \in \Dens(\Hilb)$, we obtain
\begin{align}
\lambda=1.
\end{align}
Thus $L=L'$ on $\Herm(\Hilb)^S$.

We are almost done. We will identify the components of the two normalised pairs. Fix an arbitrary state $s^{\ast}\in S$ and an arbitrary $A\in\Herm(\Hilb)$. Consider the element $Z \in \Herm(\Hilb)^{|S|}$ whose components are:
\begin{align}
Z_r=
\begin{cases}
A,&r=s^{\ast},\\
0,&r\neq s^{\ast}.
\end{cases}
\end{align}
Since $L(Z)=L'(Z)$, we have
\begin{align}
p(s^{\ast})\Tr(\rho A)
=
p'(s^{\ast})\Tr(\rho'A)
\end{align}
for every Hermitian operator $A$. In other words
\begin{align}
    \Tr[(p(s^{\ast})\rho-p'(s^{\ast})\rho')A] = 0, \ \forall A \in \Herm(\Hilb) 
\end{align}
Consequently, it follows that
\begin{align}
p(s^{\ast})\rho=p'(s^{\ast})\rho'.
\label{Eq.TracePsandA}
\end{align}
Taking the trace of the equation above, and recalling that both $\rho$ and $\rho'$ are density matrices, yields
\begin{align}
p(s^{\ast})=p'(s^{\ast}).
\end{align}
Since $s^{\ast}$ was arbitrary, this proves $p=p'$.

Finally, since $p$ is a probability distribution, there exists at least one state of the world $s_0\in S$ such that $p(s_0)>0$. Using Eq.\eqref{Eq.TracePsandA} for this particular $s_0$ and the equality $p=p'$, we have
\begin{align}
p(s_0)\rho=p(s_0)\rho'.
\end{align}
Dividing both sides by $p(s_0)$, we have proven that $\rho=\rho'$.
\end{proof}

In sum, an admissible pair $(p,\rho) \in \Delta(S) \times \Dens(\Hilb)$ can be uniquely identified by the weak ordering it induces on the entire act space. We now turn to a more local question: what can be inferred when two admissible pairs induce the same ordering on a single state-contingent slice?

\subsection{The Classical Subspace}\label{SubSec.ClassicalSubspace}

Because we are working on a presumably more general setup, one can expect that it is possible to recover the classical Anscombe-Aumann results directly from ours. In fact, as we will argue in this subsection, this is exactly the case. Basically, our representation recovers ordinary classical expected utility when quantum consequences are restricted to scalar multiples of the identity. We make this assertion more precise below. 

\begin{definition}[Scalar Acts]
    Let $\Acts$ be the family of acts. We define the \emph{scalar acts} $\Ccal$ as the convex and affinely-isomorphic to $[0,1]^S$ set as below:
\begin{align}
\Ccal=\{f_x \in \Acts; f_x(s)=x(s)I,\ x\in[0,1]^S\}.
\label{Eq.DefScalarActs}
\end{align}
    \label{Def.ScalarActs}
Additionally, for $A \subseteq S$, define the \emph{characteristic act} 
\begin{align}
e_A(s):= f_{\mathbf1_A}(s)= \begin{cases}I,&s\in A,\\0,&s\notin A.\end{cases}
\tag{11}
\label{Eq.DefCharacteristicAct}
\end{align}
\end{definition}

\begin{proposition}[Classical restriction]
Let
\begin{align}
\Pcal:= \{p \in \Delta(S); \exists \ \rho \in \Dens(\Hilb), \mbox{ with } \ (p,\rho)\in\mathcal R\}
\end{align}
be the projection of $\mathcal{R}$ onto $\Delta(S)$. Then, for every $x,y \in [0,1]^{S}$ the following statement holds true:
\begin{align}
f_{x}\succeq  f_{y}\iff p\cdot x\ge p\cdot y\quad\forall p\in\Pcal.
\label{Eq.PropositionClassicalRestriction}
\end{align}
\end{proposition}
\begin{proof}
Given $f_x\in\Ccal$, we know that
\begin{align}
U_{p,\rho}(f_x) &=\sum_s p(s)\Tr(\rho x(s)I) \nonumber \\ &=\sum_s p(s)x(s) \\
&= p \cdot x \nonumber
\end{align}
because $\Tr\rho=1$. The equivalence follows directly from the above equality.
\end{proof}

\begin{definition}[Ordering on Events]
    Let $S$ be the set of all the states of the world. Define a preorder $\succeq_{S}$ over the set of events $2^S$ as:
    \begin{align}
        A\succeq_S B \iff e_A\succeq e_B.
    \end{align}
    \label{Def.OrderOverS}
\end{definition}

We can then identify a family of probability measures that represents the preorder defined over the set of events. 

\begin{proposition}
     With the preorder over events defined above, we can show that
\begin{align}
A\succeq_S B\iff p(A)\ge p(B)\quad\forall p\in\Pcal, 
\label{Eq.PropOrderOverSOrderOverF}
\end{align}
where $p(A):=\sum_{s \in A}p(s)$.
\end{proposition}

\begin{proof}
By definition, $A\succeq_S B$ if and only if $e_A\succeq e_B$. Since $e_A=f_{\mathbf 1_A}$ and $e_B=f_{\mathbf 1_B}$, the classical-restriction proposition gives
\[
A\succeq_S B\iff p\cdot\mathbf 1_A\ge p\cdot\mathbf 1_B\quad\text{for every }p\in\Pcal.
\]
The two scalar products are $p(A)$ and $p(B)$, respectively.
\end{proof} 
 
The next lemma gives an event-mixture obstruction: a nontrivial mixture of two characteristic acts can be a characteristic act only when the two events coincide. In particular, the characteristic acts are the vertices of the scalar-act cube $\Ccal\simeq[0,1]^S$. 

\begin{lemma}[Event-mixture obstruction]
If $0<\lambda<1$, then
\begin{align}
\lambda e_A+(1-\lambda)e_B=e_C
\end{align}
if and only if $A=B=C$.
\end{lemma}
\begin{proof}
At a state in $A\setminus B$ the mixture equals $\lambda I$, and at a state in $B\setminus A$ it equals $(1-\lambda)I$. These are neither $0$ nor $I$. Hence the symmetric difference must be empty, so $A=B$. The mixture is then $e_A$, and therefore $C=A=B$.
\end{proof}

Each $e_A$ is also an extreme point of the full act space $\Acts$: every coordinate is either $0$ or $I$, and both are extreme effects. Thus the characteristic acts are extreme in both the scalar-act cube and the product act space.

Finally, we can show that it is possible to envelope the set of admissible probabilities from $\Pcal$.

\begin{proposition}
    Define the coordinate-wise probability envelopes
\begin{align}
m(s)=\inf_{p \in \Pcal}p(s),\mbox{ and }
M(s)=\sup_{p \in \Pcal}p(s).
\end{align}
Then every $p  \in \Pcal$ satisfies
\begin{align}
m(s) \leq  p(s) \leq M(s).
\end{align}
\end{proposition}

We emphasise that the functions $m$ and $M$ are only coordinate-wise lower and upper envelopes for $p$; they need not themselves be probability distributions. The envelopes summarise coordinate-wise uncertainty, but they do not in general determine all event comparisons. Two families can have the same coordinate-wise infima and suprema while disagreeing on joint restrictions across states. The full projected family $\Pcal$, rather than the envelopes alone, is therefore the fundamental classical object.

\subsection{State-Contingent Quantum Slices}\label{SubSec.StateContigentQuantumSlices}

\begin{definition}[State-Contingent Act]
    Given $s\in S$ and $E\in\Eff$, the state-contingent act $f_{s,E}$ is defined as follows 
\begin{align}
f_{s,E}: S &\to \Eff \nonumber \\
t &\mapsto f_{s,E}(t)=\begin{cases}E,&t=s,\\0,&t\ne s.\end{cases}
\end{align}
    \label{Def.StateContingency}
\end{definition}

\begin{proposition}
    Let $s$ be a particular state of the world, and $E$ an effect. Then,
    \begin{align}
U_{p,\rho}(f_{s,E})=p(s)\Tr(\rho E).
    \label{Eq.PropStateContingentFactorisation}
    \end{align}
    \label{Prop.StateContingentFactorisation}
\end{proposition}
\begin{proof}
In the defining sum for $U_{p,\rho}(f_{s,E})$, every term except the one indexed by $s$ vanishes. The remaining term is $p(s)\Tr(\rho E)$.
\end{proof}
 
In other words, state-contingent acts provide a useful diagnostic for the representation. They isolate one classical state and one quantum consequence at a time, thereby making transparent both the role of the probability weight and that of the density operator in the utility representation.

Additionally, state-contingent acts also constrain admissible pairs when they induce the same ordering on a fixed state slice. The following lemma makes this observation more concrete; its proof is the same-kernel argument used in Lemma~\ref{LemmaPositiveProportionality}. 

\begin{lemma}[Ordering identification on a state-contingent slice]
If normalized pairs $(p,\rho)$ and $(p',\rho')$ in $\Delta(S) \times \Dens(\Hilb)$ with $p(s),p'(s)>0$ induce the same ordering on all state-contingent $f_{s,E}$, then
\begin{align}
p(s)\rho=\lambda_s p'(s)\rho'
\end{align}
for some $\lambda_s>0$.
\end{lemma}
\begin{proof}
The restricted functionals are $E\mapsto\Tr(p(s)\rho E)$ and $E\mapsto\Tr(p'(s)\rho'E)$. Perturbations of $I/2$ extend their common ordering to all Hermitian directions, so the same-kernel argument used in the positive-proportionality result (lemma \ref{LemmaPositiveProportionality}) gives positive proportionality.
\end{proof}

The next proposition shows where the quantum-like weight makes a difference, independently of the classical probability.

\begin{proposition}[Support and visibility]
If $p(s)=0$, then $U_{p,\rho}(f_{s,E})=0$ for all $E$. If $p(s)>0$, then
\begin{align}
f_{s,E}\succeq_{p,\rho}f_{s,F}
\iff
\Tr(\rho E)\ge\Tr(\rho F),
\label{Eq.PropSupportVisibility}
\end{align}
where the pair-induced relation is defined by $f\succeq_{p,\rho}g$ if and only if $U_{p,\rho}(f)\ge U_{p,\rho}(g)$.
\end{proposition}

The claim follows immediately from the state-contingent slice formula: when $p(s)>0$, multiplication by $p(s)$ does not change the ordering over effects. If $p(s)=0$, this particular pair assigns zero utility to every act on that slice. For the full preference represented by $\Rmax$, a state is invisible on its quantum slice only if every pair in $\Rmax$ assigns it zero probability.





\section{Conclusion}\label{Sec.DiscussionConclusion}

If completeness is an artificial artefact in decision-theoretic frameworks---an undesirable echo of the commensurability hypothesis---is it possible to remove it from the game and retain any meaningful result? Several prior works have responded in the affirmative. This work contributes to this body of literature by showing that, by choosing the set of consequences wisely, we can not only interpret effects as physical chances but also view density operators (as well as probabilities) as subjective, probabilistic assignments.   

We did so by implementing a variation of the Anscombe--Aumann framework; that is, by imposing certain conditions on an agent's rational preferences, we could derive a representation theorem that factorises into two slices: a classical-like one and a quantum-like one. Our main representation result says that to suitably compare to acts, rational agents ought to consider affine utility functionals like the one below:
\begin{equation}
U_{p,\rho}(f)=\sum_s p(s)\Tr(\rho f(s)).
\label{Eq.ConclusionFactorisation}
\end{equation}
The probability $p$ answers the classical question ``how are the states weighted?'', whereas the density operator $\rho$ answers the quantum question ``how are effects evaluated?'' The representation therefore retains the central separation of Anscombe--Aumann theory while replacing scalar consequences by quantum effects. The role of incompleteness is equally important. A complete theory would select one ordering and, after normalisation, one pair $(p,\rho)$. Here the preference may leave several pairings admissible. 

The state-independence discussion deserves particular emphasis. Behaviourally, it is natural to require that moving the same consequences from one classical state to another should not change their aggregate comparison. Under incompleteness, however, this behavioural statement lives at the level of the intersection of supporting orderings. 
This is why we had to impose it directly at the level of the utilities to obtain the classical--quantum factorisation. Granted, in comparison with Anscombe and Aumann's framework, this is a limitation of the present representation theorem, but it is also a useful boundary: removing that condition would require a stronger selection theorem for multi-utility representations---one that we could not find and that will certainly be explored in future generalisations of this work to the setup of generalised probability theories.

The scalar-effect subspace shows that the classical part of the theory is not added after the quantum representation has been built. It is already contained in the consequence space. For scalar acts, $\rho$ disappears because $\Tr(\rho)=1$, leaving ordinary expected utility with respect to $p$. Characteristic events are extremal points of the scalar-act cube, and their comparisons yield upper and lower probability bounds. The coordinate-wise envelopes $m$ and $M$ provide lower and upper bounds on admissible probabilities. Those bounds are informative summaries, are not probabilities themselves, and should not be mistaken for a complete description of the projected family.


Before concluding, it is necessary to put this work in a historical perspective. Granted, we sought to extend Anscombe and Aumann's argument for subjective probabilities—and we did so in a quantum-like paradigm—but our rationale clearly finds counterparts elsewhere in the literature. To begin with, if we had imposed a normalisation constraint on the allowed consequences, say
\begin{align}
    \sum_{s \in S}f(s)=I
\end{align}
then Eq.\eqref{Eq.ConclusionFactorisation} is surprisingly similar to the quantity giving the probability of correctly guessing a true state out of an ensemble \cite{FD16, YKL75}—it is as if an act is preferred to another if it increases the lower bound of guessing the correct input. Along the same lines, Eq.\eqref{Eq.PropSupportVisibility} may be readily identified as a comparison involving the calculation of the Born Rule \cite{Zhang26}; what is more striking is that, in the path to obtain that particular comparison, we had to impose a sort of noncontextuality hypothesis at the level of the utility functionals. We do not want to push either of these interpretations any further, not least because this is not the intent of this paper, but also because other references have already done so in the past—with deeper analysis. In particular, by assuming a bet scheme scenario (that we do not here), the authors of \cite{BFZ17} re-derive a type of Gleason theorem. As we mentioned in the introduction, we stress that our sole intent here was to provide a formal justification for viewing quantum theory as a theory of probabilistic assignments. Finally, this contribution is not the first to champion the idea that degrees of belief can be represented by far richer objects than a single utility function or classical probabilistic assignments. Especially for the latter, it is curious that the possibility of going beyond the setup of classical probability theory had already been anticipated by several other authors working on the foundations of probability; see references \cite{Gillies10,Fishburn86,FL15} and in particular ref. \cite{Keynes1921}. We believe the present work complements that research line.

Several directions remain open. Most notably, as we mentioned before, one could seek a multi-utility representation theorem showing that an appropriate behavioural state-independence condition can be inherited by the individual supporting utilities. Such a result would remove the strongest representational assumption from the present theorem. We do not pursue that question here, but we think the most appropriate site for settling this matter is within a framework more general than quantum theory. Additionally, since suitable preference relations can give rise to subjective probability assignments, we may wonder whether agreement theorems (or the impossibility of agreeing to disagree, to be more precise \cite{LD26}) can also be formulated in this language and whether their quantum or post-quantum counterparts would finally have any difference from the classical one. To conclude, other extensions can also be addressed, in particular Savage's groundbreaking derivation and Richard Jeffrey's logic of decision; both would resolve once and for all that quantum theory is a theory of probabilistic assignments---that greater programme this paper is just a part of. 



\begin{acknowledgments}
CD thanks the hospitality of the Institute of Quantum Studies and the fellowship of Matthew Leifer, whose infinite patience was fundamental to the author's understanding of this line of research. CD also thanks the members of the Quantum Stuff Group (UFJF-Brazil); their valuable criticisms have certainly influenced the choice of topics here. Finally, a special thanks to Fernando de Melo for pushing the comparison with the Born Rule. During the development and preparation of this manuscript, ChatGPT (GPT-6, OpenAI) was used as an AI-assisted research tool to support initial mathematical brainstorming, proof checking, examination of assumptions, the search for trivial counterexamples, and editorial revision. This work was supported by CNPq through two grants from the Conhecimento Brasil Program (Lines 1 and 2).
\end{acknowledgments}

\bibliography{literature}

\end{document}